\documentclass[prd,eqsecnum,notitlepage,%linenumbers,
nofootinbib,superscriptaddress]
{revtex4-1}

\global\arraycolsep=2pt
\usepackage{stmaryrd}
\usepackage{amsmath}
\usepackage{amssymb}
\usepackage{graphicx}
\usepackage{textcomp}
\usepackage{calrsfs}
\usepackage{yfonts}
\usepackage{bm}
\usepackage{color}
\newcommand{\ben}{\begin{equation*}}
\newcommand{\een}{\end{equation*}}
\newcommand{\bean}{\begin{eqnarray*}}
\newcommand{\eean}{\end{eqnarray*}}

\newcommand{\nn}{\nonumber}
\newcommand{\be}{\begin{equation}}
\newcommand{\ee}{\end{equation}}
\newcommand{\bea}{\begin{eqnarray}}
\newcommand{\eea}{\end{eqnarray}}

\DeclareMathOperator{\Tr}{Tr}
\DeclareMathOperator{\tr}{tr}

\DeclareMathOperator{\sgn}{sgn}

\begin{document}
\title{Self-force on moving electric and magnetic dipoles:  dipole radiation,
Vavilov-\v{C}erenkov radiation,  friction with a conducting surface,
and the Einstein-Hopf effect}
\author{Kimball A. Milton}
  \email{kmilton@ou.edu}
  \affiliation{H. L. Dodge Department of Physics and Astronomy,
University of Oklahoma, Norman, OK 73019, USA}
\author{Hannah Day}
\email{Hannah.J.Day-1@ou.edu}
\affiliation{H. L. Dodge Department of Physics and Astronomy,
University of Oklahoma, Norman, OK 73019, USA}
\author{Yang Li}
\email{liyang@ou.edu}
 \affiliation{H. L. Dodge Department of Physics and Astronomy, University 
of Oklahoma, Norman, OK 73019, USA}
%\author{Prachi Parashar} 
%  \email{Prachi.Parashar@jalc.edu}
%  \affiliation{John A. Logan College, Carterville, IL
%62918, USA}
%\affiliation{Department of Energy and Process Engineering,
%Norwegian University of Science and Technology, 7491 Trondheim, Norway}
\author{Xin  Guo}
\email{guoxinmike@ou.edu}
\affiliation{H. L. Dodge Department of Physics and Astronomy, University of
Oklahoma, Norman, OK 73019, USA}
\author{Gerard Kennedy}\email{g.kennedy@soton.ac.uk}
\affiliation{School of Mathematical Sciences,
University of Southampton, Southampton, SO17 1BJ, UK}

\begin{abstract}
The classical electromagnetic self-force on an arbitrary time-dependent
 electric or magnetic dipole moving with constant velocity in vacuum, and in a 
medium, is considered. Of course, in vacuum 
 there is no net force on such a particle.  Rather, because
of loss of mass by the particle due to radiation, the self-force
precisely cancels this inertial effect, and thus the spectral 
distribution of the energy radiated
by dipole radiation is deduced without any consideration of radiation
fields or of radiation reaction, in both the nonrelativistic and
relativistic regimes.  If the particle is moving in a homogeneous medium faster
than the speed of light in the medium, 
 Vavilov-\v Cerenkov radiation results. This
is derived for the different polarization states, in agreement with
the earlier results of Frank. The friction
experienced by a point  (time-independent) dipole moving
parallel to an imperfectly conducting surface is examined. Finally,
the quantum/thermal Einstein-Hopf effect is rederived.
We obtain a closed form for the spectral distribution of the force, and
demonstrate that, even if the atom and the blackbody background have 
independent temperatures, the force is indeed a drag in the case that the
imaginary part of the polarizability is proportional to a power of the
frequency.
%for different temperatures of the atom and of the blackbody background, 
%the quantum frictional force is negative for power-law behavior
%of the imaginary part of the polarizability.
\end{abstract}

\date\today
\maketitle

\section{Introduction}
\label{sec:intro}
Quantum electrodynamic friction has a long history. It is typically
referred to as Casimir or quantum friction, and occurs when a polarizable
object or atom moves parallel to another such body.  For a review with 
many references, see Ref.~\cite{Milton:2015aba}.

But frictional drag also occurs in the classical regime.  Recently,
we considered the friction on a charged particle passing parallel to
a metallic surface described by the Drude model \cite{clfriction}.
We have now investigated the analogous effect when the particle is
neutral, but carries an  electric or a magnetic dipole moment.  We report
our results here.

In the course of this investigation, we discovered that, even if
the  metallic surface is not present, that is, if the particle is moving
with constant velocity
in vacuum, a classical finite self-force can arise. This seems rather
different from the Einstein-Hopf effect \cite{eh} due to thermal
fluctuations in the electromagnetic field, or the Boyer effect 
\cite{thb} due to zero-point fluctuations.  The  effect we observe
does not, of course, mean that the particle slows down (because the
configuration  can be obtained from that of the particle at rest
by a Lorentz transformation).  Rather,
when the dipole moment possesses time dependence, the radiation
produced by the dipole decreases the mass of the particle slightly, so
the corresponding decrease in the particle's (relativistic) momentum, 
transferred to the momentum of the radiation field,
is interpreted as  the impulse of a drag force.  
In fact, this force %, averaged over time,  
is proportional to the total energy radiated by the dipole.
That is, the formula for the energy spectrum emitted by dipole radiation, or
equivalently, the radiation reaction force, is derived entirely from
the Lorentz force law and the Maxwell-Heaviside equations without any
explicit reference to radiation fields.

If the vacuum is replaced by a homogeneous medium of permittivity 
$\varepsilon(\omega)$, and the particle moves
at a speed $v$  faster than the speed of
light in the medium $1/\sqrt{\varepsilon}$, Vavilov-\v{C}erenkov radiation
is produced.  This was worked out for electric and magnetic dipoles
many years ago by Frank \cite{frank}, although there seemed to be
some ambiguity in the case of a magnetic dipole \cite{frank84}.  
We resolve any such 
ambiguity here by doing a direct calculation using our formalism.

Then, we return to the original problem, that of a neutral particle, 
possessing either an electric or a magnetic dipole moment, moving parallel to
an imperfectly conducting surface.  Because of dissipation in the material, 
the particle experiences a frictional force, which depends on
the direction of orientation of the dipole.  Although our approach is general,
we limit our detailed results to a time-independent dipole, in
the low-velocity limit, which is
 the regime most likely to be 
 accessible to observation.

Finally, we begin our foray into the quantum vacuum regime, by quantizing our
system, replacing the product of  dipole moments 
by its fluctuational average using the
fluctuation-dissipation theorem.  Field fluctuations  are also considered.
The result is equivalent to that given by 
 Dedkov and Kyasov \cite{dedkov} and Volokitin and Persson \cite{vp},
which reduces to the Einstein-Hopf effect \cite{eh}
in the nonrelativistic limit. 

The outline of this paper is as follows.  In the next section we give
general expressions for the self-force on a moving  dipole.
In Sec.~\ref{sec3} we consider the simplest case, where an electric dipole,
moving through the vacuum,  is
oriented in the same direction as the motion, while perpendicular
orientation is treated in Sec.~\ref{sec4}.  The interpretation in terms
of dipole radiation is
given in Sec.~\ref{sec5}. In Sec.~\ref{secmag} it is shown that equivalent 
formulas are obtained for a moving magnetic dipole.  The relation between
the currents we use and the electric and magnetic polarizations are
also given there.  In Sec.~\ref{sec6} we 
consider 
Vavilov-\v{C}erenkov radiation in a uniform dielectric medium due to
a moving time-independent electric or magnetic dipole, and obtain,
using our machinery, the energy loss rate, or frictional force, 
and the corresponding energy spectrum, 
first derived by Frank in 1942 \cite{frank}, and, in particular,
confirm the second form proposed by Frank in 1984 \cite{frank84}
(and see earlier references given there).
In Sec.~\ref{sec7} we examine the classical friction experienced by  a
time-independent
dipole moving parallel to an imperfectly conducting plate, complementing
Ref.~\cite{clfriction}.  Finally, in Sec.~\ref{sec8}, we use the fluctuation-%
dissipation theorem to connect the corresponding quantum friction in
vacuum to the Einstein-Hopf effect.
Concluding remarks follow in Sec.~\ref{concl}. 
 In Appendix \ref{app0} we show how the current densities
for a moving time-dependent dipole are obtained by a Lorentz transformation 
from those in the rest frame.
%Since the Lorentz
%transformations of the currents involved may be a bit subtle, we
%supply details in Appendix \ref{app0}.
 In  Appendix \ref{appa} we give
some properties of the Green's functions used, while in Appendix \ref{appb}
we illustrate how our
method reproduces the usual static Casimir-Polder force in general.
Appendix \ref{appc} describes how the blackbody spectrum appears in a moving
frame.  Appendix \ref{appz} demonstrates that the Einstein-Hopf friction
is indeed a drag force when the imaginary part of the polarizability
is a power law in the frequency.

In this paper we use Heaviside-Lorentz (rationalized) electromagnetic
units, and set $c=\hbar=1$.

\section{Self-force on dipole}
The force is computed from the Lorentz  law for the force density,
\be
\mathbf{f}(\mathbf{r},t)=\rho(\mathbf{r},t)\mathbf{E}(\mathbf{r},t)
+\mathbf{j}(\mathbf{r},t)\times\mathbf{B}(\mathbf{r},t).\label{lorentzf}
\ee
For a time-dependent electric dipole moving with constant velocity $\mathbf{v}
=\mathbf{\hat x}v$
in the $x$ direction, the charge and current densities are
\begin{subequations}
\label{chargeandcurrent}
\bea
\rho(\mathbf{r},t)&=&-\mathbf{d}(t)\cdot\bm{\nabla}\delta(x-vt)\delta(y)
\delta(z),\\
\mathbf{j}(\mathbf{r},t)&=&-\mathbf{v}\,\mathbf{d}(t)\cdot\bm{\nabla}
\delta(x-vt)\delta(y) \delta(z)
+\mathbf{\dot d}(t)\delta(x-vt)\delta(y) \delta(z).\label{current}
\eea
\end{subequations}
Here $\mathbf{d}(t)$ is the dipole moment in the moving (lab) frame.  The
relation between the dipole moment in the moving frame and that in the rest
frame is given by (\ref{ltofdip}) below.
This current is that obtained by a Lorentz boost from the charge density
of a time-dependent dipole at rest.  (See Appendix \ref{app0}.)
The second term in the current density arises because the dipole
moment is assumed time-dependent, so it is required by current conservation:
\be
\frac{\partial}{\partial t}\rho+\bm{\nabla}\cdot \mathbf{j}=0.\label{chcon}
\ee
(Adding a curl term to $\mathbf{j}$  in the rest frame
would correspond to an intrinsic
magnetic dipole moment. See Sec.~\ref{secmag}.)
In the following, it will be convenient to use the time Fourier
transform of these quantities:
\begin{subequations}
\bea
\rho(\mathbf{r},\omega)&=&-\frac1v
\bm{\nabla}\cdot\left[\mathbf{d}\left(\frac{x}v
\right)e^{i\omega x/v}\delta(y)\delta(z)\right],\label{rhoomega}\\
\mathbf{j}(\mathbf{r},\omega)&=&-\frac{\mathbf{v}}v
\bm{\nabla}\cdot\left[\mathbf{d}\left(\frac{x}v
\right)e^{i\omega x/v}\delta(y)\delta(z)\right]+\frac{d}{dx} \mathbf{d}
\left(\frac{x}v\right)e^{i\omega x/v}\delta(y)\delta(z).\label{jomega}
\eea
\label{elcurrent}
\end{subequations}

In the frequency domain, the electric field can be expressed in terms of 
the Green's function, 
\be
\mathbf{E}(\mathbf{r};\omega)=-\frac1{i\omega}\int (d\mathbf{r}')
\bm{\Gamma}(\mathbf{r,r'};\omega)\cdot\mathbf{j}(\mathbf{r'};\omega).
\ee
It will be convenient to  adopt
%parallel to the particle's motion in the
%$x$ direction), 
the transverse Fourier representation
\be
\bm{\Gamma}(\mathbf{r,r'};\omega)=\int\frac{(d\mathbf{k_\perp})}{(2\pi)^2}
e^{i\mathbf{k_\perp\cdot(r-r')_\perp}}\mathbf{g}(z,z';\mathbf{k_\perp},
\omega),\label{Gr}
\ee
because we have in mind, as treated in 
Sec.~\ref{sec7}, 
motion next to a surface  in the $x$-$y$ plane.
This assumes the
 system has translational invariance in $x$ and $y$ directions, for which
we have the break-up into transverse electric (TE or $E$) and transverse 
magnetic (TM or $H$) modes, with $k^2=\mathbf{k}_\perp^2$, in a vacuum
region:
\be
\mathbf{g}=\left(\begin{array}{ccc}
\frac{k_x^2}{k^2}\partial_z\partial_{z'}g^H+\frac{k_y^2}{k^2}\omega^2 g^E
&\frac{k_xk_y}{k^2}\left(\partial_z\partial_{z'}g^H-\omega^2g^E\right)&
ik_x\partial_zg^H\\
\frac{k_xk_y}{k^2}\left(\partial_z\partial_{z'}g^H-\omega^2g^E\right)&
 \frac{k_y^2}{k^2}\partial_z\partial_{z'}g^H+\frac{k_x^2}{k^2}\omega^2 g^E
&ik_y\partial_{z}g^H\\
-ik_x\partial_{z'}g^H&-ik_y\partial_{z'}g^H&k^2 g^H\end{array}\right),
\label{gee}
\ee
where the transverse electric and transverse magnetic Green's functions
are equal in vacuum,
\be
g^E(z,z';\mathbf{k_\perp},\omega)=g^H(z,z';\mathbf{k_\perp},\omega)=
\frac1{2\kappa}e^{-\kappa|z-z'|},\label{gvac}
\ee
 with $\kappa=\sqrt{k_x^2+k_y^2-\omega^2}$, which is, in general,
complex. In a medium with permittivity $\varepsilon(z)$ we
replace in Eq.~(\ref{gee}) $g^H(z,z')\to g^H(z,z')/(\varepsilon(z)
\varepsilon(z'))$ with the understanding that the permittivity factors
are not acted on by the differential operators $\partial_z$, $\partial_{z'}$,
while $\kappa\to\sqrt{k^2-\omega^2\varepsilon(\omega)}$.  See Sec.~\ref{sec6}.

To compute the Lorentz force on the moving
dipole, we first eliminate, by use of Eq.~(\ref{chcon}), 
the charge density from the time averaged force density 
in Eq.~(\ref{lorentzf}), integrated over all space, 
\bea
\overline{\mathbf F}T
&=&\int\frac{d\omega}{2\pi}\int(d\mathbf{r})\left[\rho(\mathbf{r};\omega)^*
\mathbf{E}(\mathbf{r};\omega)+\mathbf{j}(\mathbf{r};\omega)^*\times
\mathbf{B}(\mathbf{r};\omega)\right]\nn\\
&=&\int\frac{d\omega}{2\pi}\int (d\mathbf{r})
(d\mathbf{r'}) \frac1{\omega^2}
j_i(\mathbf{r};\omega)^*\bm{\nabla}\Gamma_{ik}(\mathbf{r,r'}
;\omega)j_k(\mathbf{r'};\omega)\equiv \Tr \frac1{\omega^2}\mathbf{j}^*
(\bm{\nabla})\bm{\Gamma} \mathbf{j}.\label{currentcurrent}
\eea
Here, $T$ is the (large) time that the configuration exists.
This form of the mean force, including both the electric and
magnetic terms, is reminiscent of the expression used in 
quantum mechanics \cite{js}. %and is obtained by using the 
%current conservation equation (\ref{chcon}) to eliminate the charge density.
Now inserting the current densities (\ref{jomega}) into this, we
obtain for the force in the direction of motion,  
taking advantage of the $\delta$ functions,
\bea
\overline{\mathbf{F}_x} T&=&\int\frac{d\omega}{2\pi}\int dx\,dx' 
e^{-i\omega x/v}
\frac1v\left[ \mathbf{d}\left(\frac{x}v\right)\cdot\bm{\nabla}\mathbf{v}
+\mathbf{\dot d}\left(\frac{x}v\right)\right]\cdot
\frac1{\omega^2}\int\frac{(d\mathbf{k}_\perp)}{(2\pi)^2}e^{ik_x(x-x')}
ik_x\mathbf{g}(z,z';\mathbf{k}_\perp,\omega)\bigg|_{z=z'}
\nn\\
&&\quad\times
\frac1v\left[\mathbf{v}\overleftarrow{\bm{\nabla}}'\cdot
\mathbf{d}\left(\frac{x'}v\right)+\mathbf{\dot d}
\left(\frac{x'}v\right)\right]e^{i\omega x'/v}.\label{fet}
\eea
Here, we have integrated by parts, and the minus sign obtained from so
doing is incorporated in the differential operator $\overleftarrow{\bm{
\nabla}}'$. % $T$ represents the large time that the system of the uniformly
%moving dipole exists.
In the transverse directions, the gradient operators are to be interpreted 
as $\bm{\nabla}_\perp=i\mathbf{k}_\perp$, $\bm{\nabla}'_\perp
=-i\mathbf{k}_\perp$.
Similarly, the $\mathbf{\dot d}$ term can be integrated by parts, yielding
a factor $i(\omega-k_xv)$.
Now, carrying out the integrations over $x$ and $x'$ leads to the Fourier
transform of the dipole moment,
\be
\tilde{ \mathbf{d}}(\omega)=\int_{-\infty}^\infty dt\, 
e^{i\omega t}\mathbf{d}(t),
\ee
so the product of dipole moments appears as 
$\mathbf{\tilde d}(\omega-v k_x)^*\mathbf{\tilde d}(\omega-vk_x)$.
%We will discuss the $\mathbf j\times B$ contribution to the force in 
%Sec.~\ref{sec4}.

\subsection{$\mathbf{ d\parallel v}$}
\label{sec3}
Suppose first the dipole is polarized parallel to the motion, that is,
$\mathbf{d}$ 
and $\mathbf{v}$ are both in the $x$ direction.  %Then, only the electric 
%part,
%not the $\mathbf{ j\times B}$ term, 
%contributes to the drag force parallel to
% the motion. 
Then, from Eqs.~(\ref{fet}) and (\ref{gee}), 
the force in the $x$ direction is
\be
\overline{F_x^{\|}}T=\frac{1}{8\pi^3}\int_{-\infty}^\infty 
d\omega\int dk_x
dk_y |\tilde{d}(\omega-vk_x)|^2ik_x\frac{(\omega^2-k_x^2)}{2\kappa},
\ee
%where in the square brackets we have identified the contributions arising 
%from $\partial_{x'}$
%and $\partial_t$.  
The $i$ is an instruction to extract the negative of the imaginary
part.   This can only arise from $\kappa$, which becomes imaginary
when $k^2<\omega^2$.  The appropriate branch of the square root is
determined by the requirement that the singularities lie in the lower-half
$\omega$ plane, since we are dealing with the retarded propagator.
Then
\be
\omega^2>k^2:\quad \sqrt{k^2-\omega^2}=-i\sgn(\omega) \sqrt{\omega^2-k^2}.
\label{sqrtdef}
\ee
When $\omega^2-k_x^2>0$, the $k_y$ integral is
simply
\be
\int_{-\sqrt{\omega^2-k_x^2}}^{\sqrt{\omega^2-k_x^2}}\frac{d k_y}{
\sqrt{\omega^2-k_x^2-k_y^2}}=\pi,\label{sqrtint}
\ee
and we are left with
\be
\overline{F_x^{\|}}T=-\frac1{16\pi^2}\int d\omega \,dk_x 
|\tilde d(\omega-v k_x)|^2
\sgn(\omega)k_x(\omega^2-k_x^2).
\ee
When we change variables by writing $\nu=\omega-k_xv$ and $k_x=u\nu$,
we obtain
\be
\overline{F_x^{\|}} T=-\frac1{16\pi^2}\int_{-\infty}^\infty d\nu\,\nu^4
|\tilde d(\nu)|^2\int_{-1/(1+v)}^{1/(1-v)} du\,u[(1+v u)^2-u^2]
=-\frac{v}{12\pi^2}
\gamma^6\int_{-\infty}^\infty d\nu\,\nu^4 |\tilde d(\nu)|^2.
\label{parallelf}
\ee
Here $\gamma=(1-v^2)^{-1/2}$ is the usual relativistic dilation factor.
%The identical  expression, of course, is obtained if we started from the
%current expression (\ref{currentcurrent}).

%The same result follows, somewhat more simply, by starting from
%a polarization source,
%\be
%\mathbf{P}(\mathbf{r},t)=\mathbf{d}(t)\delta(x-vt)\delta(y)\delta(z),
%\ee
%together with the corresponding electric field,
%\be
%\mathbf{E}(\mathbf{r},t)=\int(d\mathbf{r'})\,dt\,
%\bm{\Gamma}(\mathbf{r,r'},t-t')\cdot\mathbf{P}(\mathbf{r'},t').
%\ee
%The force is computed from
%\be
%\overline{\mathbf{F}}T
%=\int(d\mathbf{r})(d\mathbf{r'})dt\,dt' P_i(\mathbf{r},
%t)\bm{\nabla}\Gamma_{ik}(\mathbf{r,r'},t-t')P_k(\mathbf{r'},t').
%\ee
%The computation, leading to the force (\ref{parallelf}), 
%is very similar to that detailed above.

\subsection{$\mathbf{d}\perp \mathbf{v}$}
\label{sec4}
If the polarization of the dipole is perpendicular to the motion, 
say in the $y$ direction,
%the magnetic term $\mathbf{j\times B}$ in the force density 
%must be taken into account.  Alternatively, 
we again start from 
Eq.~(\ref{fet}).
Following the same procedure we detailed in the
previous subsection, and inserting the appropriate components of the
reduced Green's function from Eq.~(\ref{gee}),  we find, after some
algebra, for the drag force in the direction of motion
\be
\overline{F^\perp_x}T=v^2 \int\frac{d\omega (d\mathbf{k_\perp})}{(2\pi)^3}
|\tilde{d}(\omega-v k_x)|^2\left(-\frac{k_x}{2i\kappa}\right)\left[
k_y^2\left(1-\frac1{v^2}\right)+\left(k_x-\frac{\omega}v\right)^2\right].
\ee
Again, carrying out the integral on $k_y$ above the branch line of the 
square root, and making the same changes of variables as before, we find
\be
\overline{F_x^\perp}T=-\frac{v^2}{16\pi^2}\int_{-\infty}^\infty d\nu\,
|\tilde d(\nu)|^2 \nu^4\int_{-1/(1+v)}^{1/(1-v)}du\,u\frac12
\left[\frac1{v^2}+\left(1+v u \left(1-\frac1{v^2}\right)\right)^2
\right]
=-\frac{v}{12\pi^2}\gamma^4\int_{-\infty}^\infty d\nu \,\nu^4 
|\tilde d(\nu)|^2.
\ee

%Again, the same result is obtained starting from 
%the polarization description, although in
%this case magnetic as well as electric polarization must be taken into
%account.  The latter is given by $\mathbf{M}'(x')=\gamma \mathbf{v}\times 
%\mathbf{P}(x)$, which is the Lorentz transform of the static electric 
%polarization $\mathbf{P}$.

\subsection{Interpretation and Discussion}
\label{sec5}
 In the above two subsections,
we have performed straightforward calculations of the classical
electromagnetic self-force on an arbitrary time-varying electric
dipole undergoing uniform motion.  The results are slightly different
depending on the orientation of the dipole.  If the dipole is 
perpendicular (parallel) to the motion, the time-averaged force in
the direction of the motion is
\be
\left.\begin{array}{c}
\overline{F_x^\perp}\\\overline{F_x^{\|}}\end{array}\right\}
=-\frac{v}{12\pi^2T}\left\{\begin{array}{c}
\gamma^4\\
\gamma^6\end{array}\right\}\int_{-\infty}^\infty d\nu\,\nu^4|\tilde d(\nu)|^2,
\label{forces}\ee
in terms of the Fourier transform of the dipole moment. 
In the above we assumed that the direction of the dipole moment was fixed.
If the direction changes with time, it is easily seen that the cross terms
cancel (because they are odd, or are proportional to $\sgn(0)=0$) 
and we have in general
\be
\overline{F_x}=-\frac{v}{12\pi^2T}\gamma^4\int_{-\infty}^\infty d\nu \,
\nu^4[ |\tilde{\mathbf{d}}_\perp(\nu)|^2+\gamma^2 |
\tilde{\mathbf{d}}_\parallel(\nu)|^2],
\ee
in terms of the dipole moment components perpendicular (parallel) to the
direction of motion.

 Three observations
immediately jump out:
\begin{itemize}
\item For an undamped oscillator of frequency $\omega_0$, the integral
over $\nu$ is proportional to $T=2\pi\delta(0)$, because $\tilde d(\nu)\propto 
\delta(\nu-\omega_0)$. 
\item In the nonrelativistic limit the drag forces are identical.
\item The friction is proportional to the total
energy radiated by an oscillating
dipole at rest \cite{ce}
\be
E'_R=\frac1{12\pi^2}\int_{-\infty}^\infty d\nu\,\nu^4|\tilde d'(\nu)|^2,
\ee
where the integrand is the spectral energy distribution of a radiating
dipole. (Primes denote quantities in the particle's rest frame.)
\end{itemize}
Thus, for low velocities, the force satisfies
\be
\overline{F_x}T=-v E_R.
\ee
This result has an extremely simple interpretation.  Of course, there
is no force on a uniformly moving dipole  in vacuum,
 since it may be obtained from
a dipole at rest by a Lorentz boost.  But, because the dipole loses 
energy $E_R$ over the course of its motion, its mass decreases accordingly,
and therefore, nonrelativistically, 
its momentum decreases by $E_R v$. This is the negative of the momentum
carried off by the radiation field.  Momentum conservation means that a
radiation reaction force  $\overline{F_x}=-E_R v/T$ 
is acting on the moving dipole.

In the relativistic regime, the different factors of $\gamma$ are easily
understood.  One factor of $\gamma$ comes from the Lorentz transformation
of the momentum of the radiated energy, $(P_R)_x=v\gamma E'_R$. 
 Then if we transform the dipole moments in the
moving frame to those in the rest frame,
\be
\gamma d_x(\gamma t)= d'_x(t),\quad d_y(\gamma t)=d'_y(t),\label{ltofdip}
\ee
the Fourier transforms are related by
\be
\tilde d_x\left(\frac{\nu}\gamma\right)=\tilde d_x'(\nu),\quad
\frac1\gamma\tilde d_y\left(\frac{\nu}\gamma\right)=\tilde d_y'(\nu).
\ee
Thus when $P_R$ is expressed in terms of the dipole moment in the
moving frame $\mathbf{d}$, 
we obtain exactly the structure seen in Eq.~(\ref{forces}).
In terms of the dipole moment in the rest frame of the particle,  the average
force is given by
\be
\overline{F_x}=-\frac{v\gamma}{12\pi^2T}\int_{-\infty}^\infty d\nu\,
\nu^4|\mathbf{\tilde d}'(\nu)|^2.\label{diprf}
\ee

\subsection{Magnetic Dipole}
\label{secmag}

What about a magnetic dipole moving through vacuum?  Precisely the same
considerations apply, and because the results are obtained by a duality
transformation, $\mathbf{d}\to\bm{\mu}$, E $\leftrightarrow$ H, and $g^H=g^E$
in vacuum,\footnote{In dual electrodynamics, the magnetic current would
have the same form as Eq.~(\ref{current}), except $\mathbf{d}\to\bm{\mu}$.}
 the same form of the inertial effect emerges:
\be
\overline{F_x}=-\frac{v}{12\pi^2T}\gamma^4\int_{-\infty}^\infty d\nu \,
\nu^4[ |\tilde{\bm{\mu}}_\perp(\nu)|^2+\gamma^2 |
\tilde{\bm{\mu}}_\parallel(\nu)|^2]=
-\frac{v\gamma}{12\pi^2T}\int_{-\infty}^\infty d\nu\,
\nu^4|\bm{\tilde \mu}'(\nu)|^2,%\label{mdiprf}
\label{magrad}
\ee
again expressed in terms of the energy loss due to magnetic dipole radiation,
where $\bm{\tilde \mu}'$ is the dipole moment in the rest frame of the
particle.

For the following applications, however, duality will fail.  So we must
use the electric current density of a moving time-dependent  magnetic dipole,  
obtained from the current of a magnetic dipole at rest
\be
\mathbf{j}'(\mathbf{r}',t')=\bm{\nabla}'\times\bm{\mu}'(t')\delta(\mathbf{r'}),
\ee
which after a boost of velocity $\mathbf{v}$ yields
\begin{subequations}
\label{magchcur}
\bea
\rho(\mathbf{r},t)&=&-[\mathbf{v}\times\bm{\mu}(t)]\cdot\bm{\nabla}\delta(
\mathbf{r}-\mathbf{v}t),\\
\mathbf{j}(\mathbf{r},t)&=&-\bm{\mu}(t)\times\bm{\nabla}\delta(\mathbf{r}
-\mathbf{v}t)+\partial_t\left[\mathbf{v}\times\bm{\mu}(t)\delta(\mathbf{r}
-\mathbf{v}t)\right],\label{magcurrent}
\eea
\end{subequations}
where $\bm{\mu}(t)$ is the magnetic dipole moment in the moving frame.
The time derivative term is required by current conservation, 
Eq.~(\ref{chcon}).
It is then straightforward algebra to show this current, inserted into
our general construction (\ref{currentcurrent}) yields the radiation
formula (\ref{magrad}), as asserted.

It actually might seem more natural to pose this problem in
terms of electric and magnetic polarizations $\mathbf{P}$ and $\mathbf{M}$, 
instead of electric currents and charges. Of course, the two are related by
\be
\rho=-\bm{\nabla}\cdot\mathbf{P},\quad \mathbf{j}=\bm{\nabla}\times \mathbf{M}
+\frac\partial{\partial t}\mathbf{P}.
\ee
By inspection of Eqs.~(\ref{chargeandcurrent}) and (\ref{magchcur}), 
it is immediately seen that
\begin{subequations}
\bea
\mathbf{P}(\mathbf{r},t)&=&
\left[\mathbf{d}(t)-\bm{\mu}(t)\times\mathbf{v}\right]\delta(\mathbf{r-v}t),\\
\mathbf{M}(\mathbf{r},t)&=&
\left[\bm{\mu}(t)+\mathbf{d}(t)\times\mathbf{v}\right]\delta(\mathbf{r-v}t).
\eea
\end{subequations}

\section{Vavilov-\v Cerenkov radiation}
\label{sec6}

The above calculations were  performed assuming the background was vacuum.  
It is easy to
extend them to the motion of a dipole through a homogeneous medium.
In this case, a new phenomenon can emerge, when the velocity of the
particle exceeds the speed of light in the medium $1/n$, where the
index of refraction is $n(\omega)=\sqrt{\varepsilon(\omega)}$.  
This is the famous
Vavilov-\v Cerenkov effect, usually considered for a charged particle
\cite{cherenkov,tamm-frank}.

We repeat the above calculations with the substitutions of the permittivity
mentioned following Eq.~(\ref{gvac}), and with 
\be
g^E(z,z')=\frac1{2\kappa}e^{-\kappa|z-z'|},\quad
g^H(z,z')=\frac\varepsilon{2\kappa}e^{-\kappa|z-z'|},\quad
\kappa=\sqrt{k^2-\omega^2\varepsilon(\omega)}.
\ee
Then, 
\be
g_{xx}(0,0)=\frac1{2\kappa}\left[\omega^2-\frac{k_x^2}{\varepsilon(\omega)}
\right],
\ee
and for a longitudinally polarized electric 
dipole we have the drag force given by
\be
\overline{F^\parallel_x}T=\frac1{16\pi^2}\int_{-\infty}^\infty d\nu\,\nu^3
|\nu||\tilde d(\nu)|^2\int du\,u\sgn(\nu(1+vu))\left[(1+vu)^2
-\frac{u^2}{\varepsilon(\nu(1+vu))}\right].
\label{fparac}
\ee
Here, the limits of the $u$ integration are determined by
\be
 \omega^2\varepsilon(\omega)>k_x^2,\quad \mbox{or}\quad
(1+vu)^2\varepsilon(\nu(1+vu))>u^2.
\ee

In the case that $\varepsilon$ is independent of frequency, the limits
become 
\be
-\frac1{1/n+v},\quad \frac1{1/n-v}, 
\ee
where $n=\sqrt{\varepsilon}$,
and then if the speed of the dipole is smaller than the speed of light
in the medium, $v<1/n$,
\be
\overline{F^\parallel_x}T=-v\frac{n^3}{(1-v^2n^2)^3} E_R,\quad
E_R=\frac1{12\pi^2}\int_{-\infty}^\infty d\nu\,\nu^4|\mathbf{\tilde d}(\nu)|^2,
\ee
in terms of the radiated dipole energy $E_R$. 
Similarly, for the transverse polarization, we find
\be
\overline{F_x^\perp}T=-v\frac{n^3}{(1-v^2n^2)^2}E_R.
\ee
 These are simple generalizations
of Eq.~(\ref{forces}).

The assumption of dispersionless permittivity is quite unrealistic, of course.
Instead, assume dispersion, but in order 
 to obtain a simple result,  let us consider only
 the case where the dipole has no
time dependence, $\mathbf{d}(t)=\mathbf{d}_0$, so there is no radiation in 
the rest frame.  But now there is Vavilov-\v Cerenkov radiation in the
frame where the dipole moves with constant velocity greater than that of
the speed of light, $v>1/n(\omega)$.  The $\nu$ integral 
in Eq.~(\ref{fparac}) becomes trivial
because
\be
\tilde{\mathbf{d}}(\omega)=2\pi\delta(\omega)\mathbf{d}_0,
\ee
where, again,  we interpret $2\pi\delta(0)=T$.  
The result for the drag in the case of a moving electric
dipole polarized in the direction parallel to the motion
is, after changing variable, $k_xv=\omega$,
\be
(F^{\parallel}_x)_d
=-\frac{d_0^2}{4\pi}\frac1{v^2}\int_{\omega>0} d\omega\,
\omega^3\left[1-\frac1{v^2n(\omega)^2}\right],\label{dpara}
\ee
where the integration extends over those  positive frequencies for
which the spectral distribution is positive.
This distribution coincides with the result of Frank \cite{frank}, Eq.~(2.33) 
there.  (See also Refs.~\cite{frank84,leonhardt19}.)

The result for perpendicular polarization is somewhat different, but
obtained in precisely the same way:
\be
(F_x^\perp)_d=-\frac{d_0^2}{8\pi}\int_{\omega>0}
 d\omega\, \omega^3n(\omega)^2\left[1-\frac1{v^2n(\omega)^2}\right]^2,
\label{dperp}
\ee
coinciding with the spectral distribution found by Frank \cite{frank}, 
Eq.~(2.34).

As noted above, in vacuum there is no difference between the formulas
for the drag forces for a magnetic dipole as compared to an electric dipole.
But this is not the case in the medium.  Nevertheless, 
the steps are just the same,
with the duality transformations, $d\to\mu$, $E\leftrightarrow H$, 
$\varepsilon\leftrightarrow\mu=1$.  So for a constant magnetic dipole
polarized longitudinally,
\be
(F_x^\parallel)_\mu=-\frac{\mu_0^2}{4\pi v^2}
\int_{\omega>0}d\omega
\,\omega^3 n(\omega)^2\left[1-\frac1{v^2n(\omega)^2}\right],\label{paracmag}
\ee
which agrees with Ref.~\cite{frank84}, Eq.~(4.35), 
but there seems to be a missing $n^2$ 
in Ref.~\cite{leonhardt19}.  For perpendicular polarization, the
result, after a bit of algebra, is
\be
(F_x^\perp)_\mu
=-\frac{\mu_0^2}{8\pi}\int_{\omega>0}d\omega\,\omega^3
n(\omega)^4\left[1-\frac1{v^2  n(\omega)^2}\right]^2,\label{perpwrong}
\ee
which agrees with one of the two alternatives, Eq.~(4.37), 
given by Ref.~\cite{frank84}.\footnote{It might be noted that Eq.~(\ref{dpara})
is obtained from Eq.~(\ref{dperp}) 
by differentiating with respect to $\ln v^2$.
The same is true for Eqs.~(\ref{paracmag}) and (\ref{perpwrong}).}

However, this treatment is suspect, because the dielectric medium breaks
dual symmetry.  We should repeat the calculation using the electric current
(\ref{magcurrent}).  Perhaps not surprisingly, the same result (\ref{paracmag})
is obtained for parallel polarization.  But that is not the case for 
perpendicular polarization.  Since this is a bit more complicated, and perhaps
controversial, let us supply a few more details.  Assuming $\bm{\mu}$ points
in the $y$-direction, with the velocity in the $x$-direction, the general
formula reduces to
\bea
(\overline{F_x^\perp})_\mu T&=&\int\frac{d\omega\,dk_x\,dk_y}{(2\pi)^3}
\frac{ik_x}{\omega^2}|\tilde{\mu}(\omega-vk_x)|^2\left[\partial_z\partial_{z'}
g_{xx}+i(k_x-v\omega)(\partial_z g_{xz}-\partial_{z'}g_{zx})+(k_x-v\omega)^2
g_{zz}\right]\nn\\
&=&T\frac{\mu_0^2}{8\pi^2}\int dk_x\,dk_y\frac{ik_x}\kappa\left[(v k_x)^2
\left(\sqrt{\varepsilon}-\frac1{\sqrt{\varepsilon}}\right)^2+k_y^2\left(
\frac{v^2}\varepsilon-1\right)\right].
\eea
Here, we have assumed the dipole has no time variation, $\bm{\mu}(t)=
\mathbf{\hat y}\mu_0$.
Carrying out the $k_y$ integration as before, we obtain the result (letting
$vk_x=\omega$)
\be
(F_x^\perp)_\mu
=-\frac{\mu_0^2}{8\pi v^2}\int_{\omega>0} d\omega\,\omega^3 n^2(\omega)
\left[2\left(1-\frac1{n(\omega)^2}\right)^2-\left(1-\frac{v^2}{n(\omega)^2}
\right)\left(1-\frac1{v^2n(\omega)^2}\right)\right].
\ee
The integral is over  those positive frequencies for which
$n(\omega)v>1$.  The integrand is
the spectral distribution of Vavilov-\v{C}erenkov
radiation found by Frank in Ref.~\cite{frank84}, Eq.~(4.36). 
It is somewhat surprising
because it is discontinuous at threshold;  that is, it is zero if $vn<1$, but
it jumps to a finite value when the particle exceeds the speed of light for
a given frequency.  The earlier result (\ref{perpwrong}), obtained by the 
plausible duality argument, is indeed incorrect.
    
The dipolar Vavilov-\v{C}enenkov effect is rather small.  Compared to
the corresponding charged particle effect, as described, for example, in
Ref.~\cite{clfriction,ce}, where the corresponding frictional force is
\be
(F_x)_e=-\frac{e^2}{4\pi}\int d\omega\,
\omega\left(1-\frac1{v^2n(\omega)^2}\right),
\ee
the electric dipole effect
 is smaller by a factor of order 
$(d/(e\lambda))^2$, where $\lambda$ is the characteristic wavelength of
the radiation emitted.  This is typically a very small number, because the
size of the particle is small compared to an optical wavelength.

\section{Friction of dipole passing close to conducting surface}
\label{sec7}
Next, we consider either an electric or a magnetic dipole moving with
constant velocity parallel to an imperfectly conducting surface
in the $x$-$y$ plane.  The idea
is an immediate generalization of the analysis of the same situation for a
charged particle \cite{clfriction}.  We start with Eq.~(\ref{fet}),
which for a time-independent electric dipole $\mathbf{d}(t)=\mathbf{d}_0$ 
yields
\be
(F_x)_d
=\frac{d_0^2v^2}{4\pi^2}\int dk_x\,dk_y \frac{ik_x}{\omega^2}\nabla_n
\nabla'_n g_{xx},\quad \omega=vk_x,
\ee
where $\nabla_n=\mathbf{\hat n}\cdot\bm{\nabla}$, with $\mathbf{\hat n}=
\mathbf{d}_0/d_0$.  Here $g_{xx}$ is given in Eq.~(\ref{gee}), but now
\be
g^{E,H}=\frac1{2\kappa}\left[e^{-\kappa|z-z'|}+r^{E,H}e^{-\kappa(z+z')}\right],
 \quad z,z'>0,
\ee
in the vacuum region above the conductor, occupying the semispace $z<0$,
where the reflection coefficients are
\be
r^E=\frac{\kappa-\kappa'}{\kappa+\kappa'},\quad r^H=
\frac{\kappa-\kappa'/\varepsilon}{\kappa+\kappa'/\varepsilon},\quad
\kappa=\sqrt{k^2-\omega^2},\quad \kappa'=\sqrt{k^2-\omega^2\varepsilon}.
\ee
The required imaginary part can only come from the reflection coefficients,
because $\kappa$ is real.
It is convenient to write the above integral in polar coordinates, with
$k_x=\gamma \kappa \cos\theta$, $k_y=\kappa\sin\theta$.

Depending on the polarization of  the dipole relative to the surface and
the direction of motion,
the force can be written in terms of transverse electric and
transverse magnetic contributions:
\be
(F_x)_d=-\frac{d_0^2}{8\pi^2}(\gamma^2-1)\int_0^\infty d\kappa\,
\kappa^3 e^{-2\kappa a}\int_0^{2\pi} d\theta
\frac{\cos\theta}{1+(\gamma^2-1)\cos^2\theta}(f^E+f^H)\left\{ 
\begin{array}{cc}
\gamma^2\cos^2\theta,&\mathbf{d_0}=\mathbf{\hat x}d_0,\\
\sin^2\theta,&\mathbf{d_0}=\mathbf{\hat y}d_0,\\
1,&\mathbf{d_0}=\mathbf{\hat z}d_0,
\end{array}\right.
\label{fxpara}
\ee
where $a$ is the distance between the trajectory of the dipole and the surface.
Here $f^{E,H}$ are the same functions appearing in Ref.~\cite{clfriction}:
\begin{subequations}
\bea
f^E(\kappa,\theta,\gamma)&=&2\sin^2\theta\Im\left[1+\sqrt{1-(\gamma^2-1)
(\varepsilon-1)\cos^2\theta}\right]^{-1}\\
f^H(\kappa,\theta,\gamma)&=&2\frac{\gamma^2}{\gamma^2-1}
\Im\left[1+\frac1\varepsilon\sqrt{1-(\gamma^2-1)
(\varepsilon-1)\cos^2\theta}\right]^{-1}.
\eea
\end{subequations}
We model the conductor by the Drude model:
\be
\varepsilon(\omega)=1-\frac{\omega_p^2}{\omega^2+i\nu\omega},\label{drude} 
\ee
in terms of the plasma frequency $\omega_p$ and the damping parameter $\nu$.

Because the $f$'s are the same as discussed previously, we can carry over
various limits from Ref.~\cite{clfriction}.  We will content ourselves here
with the low-velocity limit, $v\ll \nu a\ll 1$.  Defining
$\hat\alpha=2\omega_p a$, $\hat\beta=2\nu a$, and $\hat u=2\kappa a$, we have 
\cite{clfriction}
\be
f^E=\frac{\hat\alpha^2 v}{4\hat u\hat\beta}\cos\theta\sin^2\theta,\quad
f^H=\frac{2\hat u\hat\beta}{\hat\alpha^2 v}\cos\theta,\quad v\ll\hat\beta\ll1.
\label{smallvf}
\ee
Obviously, the TE contribution is subdominant, and the leading behavior is
\be
(F^{x,y,z}_x)_d\approx  -\frac{3 d_0^2}{32\pi a^4}\frac{\hat\beta v}
{\hat\alpha^2}
\left(3,1,4\right),
\quad v\ll\hat\beta\ll 1,
\label{fepara}
\ee
where the superscript refers to the direction of polarization of 
$\mathbf{d}_0$.

%When the dipole is oriented perpendicular to the motion, but
%parallel to the plane of the conductor,  $\mathbf{d}=d_0\mathbf{\hat y}$, 
%%in place of Eq.~(\ref{fepara}), we have
%the force is given by a formula of the same form as Eq.~(\ref{fxpara}) but 
%with
%\be
%(F^y_x)_d=-\frac{d_0^2}{8\pi^2}(\gamma^2-1)\int_0^\infty d\kappa\,
%\kappa^3 e^{-2\kappa a}\int_0^{2\pi} d\theta\,\frac{\cos\theta\sin^2\theta} 
%G^y_d=\sin^2\theta(f^E+f^H).
%\label{fxperp}
%\ee
%Again, inserting the dominant expression for $f^H$, we obtain for low
%velocities,
%\be
%(F_x^y)_d=-\frac{3 d^2_0}{32\pi a^4}\frac{\beta v}{\alpha^2},
%\quad v\ll\beta\ll 1.
%\ee

%Finally, we look at the dipole being oriented perpendicular to both the motion
%and the surface, $\mathbf{d}=d_0\mathbf{\hat z}$.  Then
% \be
%(F^z_x)_d=-\frac{d_0^2}{8\pi^2}(\gamma^2-1)\int_0^\infty d\kappa\,
%\kappa^3 e^{-2\kappa a}\int_0^{2\pi} d\theta\,\frac{\cos\theta} 
%{1+(\gamma^2-1)\cos^2\theta}
%G^z_d=f^E+f^H,
%\label{fxperp2}
%\ee
%which yields
%\be
%(F_x^z)_d=-\frac{3 d_0^2}{32\pi a^4}\frac{4\beta v}{\alpha^2}, 
%\quad v\ll\beta\ll 1.
%\ee
%Note these three contributions are in the ratio $3:1:4$.

The magnetic friction is derived similarly.  We use the current 
(\ref{magcurrent}) to write the force as ($\omega=vk_x$)
\be
(F_x)_\mu=\int\frac{dk_x\,dk_y}{(2\pi)^2}\frac{ik_x}{\omega^2}\bm{\mu}_0\times
\left(\bm{\nabla}-i\mathbf{v}\omega\right)\cdot\mathbf{g}\cdot\bm{\mu}_0
\times\left(\overleftarrow{\bm{\nabla}'}+i\mathbf{v}\omega\right).
\ee
For the different orientations of the magnetic dipole, we
carry out  the algebra and find a result of the form of Eq.~(\ref{fxpara}),
or
\be
(F_x)_\mu=-\frac{\mu_0^2}{128\pi^2a^4}(\gamma^2-1)\int_0^\infty
d\hat u\,\hat u^3e^{-\hat u}\int_0^{2\pi}d\theta\frac{\cos\theta}{1+(\gamma^2-1)\cos^2\theta}
(\tilde{f}^E+\tilde{f}^H)\left\{
\begin{array}{cc}
\gamma^2\cos^2\theta,&\bm{\mu}_0=\mathbf{\hat x}\mu_0,\\
\sin^2\theta,&\bm{\mu}_0=\mathbf{\hat y}\mu_0,\\
1,&\bm{\mu}_0=\mathbf{\hat z}\mu_0,
\end{array}\right.,
\ee
 which looks just like Eq.~(\ref{fxpara}) except that 
$\mathbf{d}_0\to\bm{\mu}_0$, and the $f$'s are replaced by
 $\tilde f$'s which  are defined by
\be
\tilde{f}^E=\frac{\gamma^2}{\gamma^2-1}\frac1{\sin^2\theta}f^E,\quad
\tilde{f}^H=\sin^2\theta \frac{\gamma^2-1}{\gamma^2}f^H.
\ee 
Now, because of the kinematic factors, the TE contribution dominates, and
using the previously stated behaviors (\ref{smallvf}) for small $v$, we
find
\be
(F_x^{x,y,z})_\mu=-\frac{\mu_0^2}{128\pi a^4}\frac{\hat\alpha^2}{8\hat\beta}v
\left(3,1,4\right),\quad
v\ll\hat\beta\ll1,
\ee
since the angular integrals are the same as for the electric dipole.

%The forces for the two other polarizations are found similarly.  In the
%low-velocity realm,
%\bea
%(F_x^y)_\mu&=&-\frac{\mu_0^2}{128\pi a^4}\frac{\alpha^2}{8\beta}v,\\
%(F_x^z)_\mu&=&-\frac{\mu_0^2}{128\pi a^4}\frac{\alpha^2}{2\beta}v,
%\eea
%where the ratios between the different polarization contributions 
%are again in the ratio $3:1:4$, 
%because the $\theta$ integrations are the same.
%\end{subequations}

Again the electric dipole friction is relatively smaller than the corresponding
electric charge case \cite{clfriction}, by a factor of order $(d/ea)^2$.

\section{Quantum Vacuum Friction}
\label{sec8}

\subsection{$dd$ Fluctuations}
Let us return to the situation of a particle moving uniformly in
vacuum, but now
quantize the dipole. We assume the particle has no mean dipole moment in
its rest frame,
$\langle \mathbf{d'}(t')\rangle=0$.
 The formula we obtained in Sec.~\ref{sec5} for the
force in the moving frame of the particle expressed in terms of the rest-frame
dipole moments, Eq.~(\ref{diprf}),
 is quantized by using the fluctuation-dissipation theorem \cite{ford2,kubo}
\be
\left\langle \frac{\mathbf{d}'(t'_1)\mathbf{d}'(t'_2)+\mathbf{d}'(t'_2)
\mathbf{d}'(t'_1)}2\right\rangle\equiv\langle\mathcal{S}\mathbf{d}'(t_1')
\mathbf{d}'(t_2')\rangle
=\int_{-\infty}^\infty \frac{d\nu}{2\pi}
e^{-i\nu(t_1'-t_2')}\Im\bm{\alpha}(\nu)\coth\frac{\beta'\nu}2,\label{ddfdt}
\ee
where we have symmetrized the dipole operators.  Here, $\beta'$ is the inverse
temperature in the particle's rest frame, and  $\bm{\alpha}$ is the
electric polarizability tensor of the particle in its rest frame.
We Fourier transform Eq.~(\ref{ddfdt}) and find the proper replacement to be 
made in Eq.~(\ref{diprf}) when quantizing the dipole is 
\be
|\mathbf{\tilde d'}(\omega)|^{2} \to T' \Im\bm{\alpha}(\omega)
\coth\frac{\beta'\omega}{2}.
\ee Here, $T'$ is the total time that the configuration exists measured in 
the rest frame of the particle, related to $T$ by  $T=\gamma T'$, because
the change in the particle coordinate is $dx'=0$.
%When we take the frequency transform of the dipole correlation function,
%we encounter the total time of the configuration in the rest frame of the
%particle, $T'=\gamma T$.
For an isotropic particle, $\bm{\alpha}=\alpha\bm{1}$,
then, the force arising from the dipole fluctuations
is
\be
(F_x)_{dd}=-\frac{v}{4\pi^2}\int_{-\infty}^\infty d\omega\,\omega^4
\Im\alpha(\omega)\coth\frac{\beta'\omega}2.\label{fddfin}
\ee

 The above argument is somewhat heuristic. For a more rigorous approach, 
consider first the free energy in the rest frame of the particle, which may be 
written as
\begin{equation}
\mathcal{F}' = -\left.\int_{-\infty}^{\infty}dt'_2 \int_{-\infty}^{\infty}
\frac{d\omega}{2\pi} \,e^{-i\omega (t'_1-t'_2)} \int
\frac{(d\bold{k}'_{\perp})}{(2\pi)^2} \,e^{i\bold{k}'_{\perp}\bold{\cdot}(\bold{r}'_1-\bold{r}'_2)_{\perp}} \tr \bold{g}'_R(z'_1, z'_2; \bold{k}'_{\perp}, 
\omega)\, \langle \mathcal{S} \bold{d}'(t'_2) \bold{d}'(t'_1)\rangle 
\right|_{\bold{r}'_{1\perp}=\bold{r}'_{2\perp}, z'_1=z'_2}.\label{fex}
\end{equation}
In the frame in which the particle is moving with velocity 
$\bold{v} =v \bold{\hat x}$, the transformed free energy is $\mathcal{F}
=\frac{1}{\gamma}\mathcal{F}'$.\footnote{Note that the action 
$W=-\mathcal{F} T$ is Lorentz invariant, which is
why $T$ and $\mathcal{F}$ transform contravariantly.} 
The corresponding force on the 
particle may then be obtained %from $(F_x)_{dd}=-\partial_{x} 
%\mathcal{F}=-\left(\partial_{x'}-v\partial_{t'}\right) \mathcal{F}'$, which 
%amounts to 
by the insertion of a factor of $-i(k'_x+v\omega)$ in 
Eq.~(\ref{fex}). Using Eq.~(\ref{ddfdt}), we then have, in general, 
%
%
%Because the above argument might seem a bit ambiguous, we repeat the
%calculation by considering the (time-averaged) free energy in the rest frame 
%of the particle, which is immediately seen to be
%\be
%\overline{\mathcal{F}'}=-\frac12 \int_{-\infty}^\infty d(t_1'-t_2') 
%\langle\mathbf{d}'(t'_1)\cdot \int_{-\infty}^\infty\frac{d\omega}{2\pi}
%e^{-i\omega(t_1'-t_2')}\int\frac{(d\mathbf{k}'_\perp)}{(2\pi)^2}
%e^{i\mathbf{k'_\perp\cdot(r'_1-r'_2)_\perp}}\mathbf{g}'(z'_1,z'_2)\cdot 
%\mathbf{d}'(t_2')\rangle
%\bigg|_{\mathbf{r'_{1\perp}=r'_{2\perp}},z'_1=z'_2}.
%\label{fex}
%\ee
%Now under a boost by $-v \mathbf{\hat x}$, 
%so the particle is moving in the $x$ direction with velocity $v$, 
%the free energy transforms to $\mathcal{F}=\frac1\gamma\mathcal{F}'$.
%\footnote{}
%To compute the force in the moving frame, we take the negative derivative 
%of the free energy in Eq.~(\ref{fex}) with respect to the moving frame 
%coordinate $x=\gamma (x'+vt')$, 
%which amounts to the insertion of a factor of $-i\gamma(k'_{x}+v \omega)$ 
%and multiplying  by a factor of two.   Then using  Eq.~(\ref{ddfdt}), we have
%in general,
\be
(F_x)_{dd}=-\int_{-\infty}^\infty\frac{d\omega}{2\pi}\tr \Im\bm{\alpha}(\omega)
\int\frac{(d\mathbf{k'_\perp})}{(2\pi)^2} (k'_x+v\omega)\Im
\mathbf{g}_R'(0,0;\mathbf{k'_\perp},\omega)
\coth\frac{\beta'\omega}2.\label{gendd}
\ee  The imaginary part of the Green's function emerges from the
symmetry under $\omega\to-\omega$, $k'_x\to-k'_x$.
For an isotropic particle, 
the trace of $\mathbf{g}(0,0)$ is in vacuum $\omega^2/\kappa$,\footnote{
The vacuum and the vacuum retarded Green's function are Lorentz invariant.
On the other hand, if we thought of $\Gamma_R$ in terms of an expectation
value of field products, those transform as described in the next subsection,
with the same result.}  where $\kappa=\sqrt{k_\perp^{\prime2}-\omega^2}$.
%
%&\to&-\frac{T'}2\int_{-\infty}^\infty \frac{d\omega}{2\pi}
%\Tr\Im\bm{\alpha}(\omega)\coth\frac{\beta'\omega}2
%\int\frac{(d\mathbf{k}_\perp)}{(2\pi)^2}\cdot
%e^{i\mathbf{k_\perp\cdot(r'_1-r'_2)_\perp}}
%\mathbf{g}(z'_1,z'_2)\bigg|_{\mathbf{r'_{1\perp}=r'_{2\perp}},z'_1=z'_2},
%\label{uinrf}
%\eea
%%
%
%The
%integral over $\mathbf{k_\perp}$ in Eq.~(\ref{uinrf}) is divergent; we regulate
%it by retaining point splitting in $x'_1-x'_2$ and in $t'_1-t'_2$.  
%% so
%\be
%\mathcal{F}=-\frac\gamma{16\pi^3}\int_{-\infty}^\infty d\omega\,\omega^2
%\Im\alpha(\omega)\coth\frac{\beta'\omega}2
%\int (d\mathbf{k_\perp}) \frac1\kappa e^{ik_x(x'_1-x'_2)}.\label{fe2}
%\ee
%
This yields
%Now to compute the force in the moving frame, we differentiate with 
%respect to $-x_1$ which amounts to the insertion of the factor of
%$-i\gamma(k_x+\omega v)$, yielding, after multiplying by two,
\be
(F_x)_{dd}=-\frac1{8\pi^3}\int_{-\infty}^\infty d\omega\,\omega^2
\Im\alpha(\omega)\coth\frac{\beta'\omega}2\int dk'_x\,dk'_y \Im\frac1\kappa
(k'_x+v\omega).
\ee
The imaginary part of $1/\kappa$ arises from the branch line in 
$\kappa$, according
to Eqs.~(\ref{sqrtdef}) and (\ref{sqrtint}).  And then doing the $k_x$ 
integration over the interval from $-|\omega|$ to $|\omega|$ we are left
with precisely Eq.~(\ref{fddfin}).

\subsection{$EE$ Fluctuations}

This is not the end of the story.
We must now also include fluctuations in the electromagnetic field.
This contribution arises from the force term
\be
(F_x)_{EE}=\frac1\gamma\int_{-\infty}^\infty \frac{d\omega}{2\pi}
\tr\bm{\alpha}(\omega)^* 
\nabla_x\langle \mathbf{E' E'}\rangle(\omega),\label{fee}
\ee 
which represents the interaction between the dipoles induced by the field
fluctuations.  Here the factor of $1/\gamma$ is present to transform 
the energy in the rest frame of the particle to that in the rest frame
of the blackbody radiation. %Although there are serious 
%ordering ambiguities which we will discuss elsewhere (and see Appendix), 
We will  here assume once again that the field operators are
to be merely symmetrized,
\be
\langle \mathbf{ E' E'}\rangle(\mathbf{r'_1,r'_2};\omega)
\equiv\int_{-\infty}^\infty d(t_1'-t_2')\,e^{i\omega(t_1'-t_2')}
\left\langle\mathcal{S}\mathbf{E'}(\mathbf{r}'_1,t'_1)
\mathbf{E'}(\mathbf{r}'_2,t'_2)\right\rangle.\label{ftee}
\ee
Here $\mathbf{E'}(\mathbf{r'},t')$ is the electric field in the rest frame
of the particle.  We need to Lorentz transform the fields to the rest frame
of the blackbody radiation in the vacuum, accomplished by using
\bea
E'_x(\mathbf{r'},t')&=&E_x(\mathbf{r},t),\nn\\
E'_y(\mathbf{r'},t')&=&\gamma(E_y(\mathbf{r},t)-v B_z(\mathbf{r},t)),\nn\\
E'_z(\mathbf{r'},t')&=&\gamma(E_z(\mathbf{r},t)+v B_y(\mathbf{r},t)),
\eea
where
\be
x'=\gamma(x-vt),\quad y'=y, \quad z'=z,\quad t'=\gamma(t-vx).
\ee  Using Maxwell's equations to eliminate $\mathbf{B}$ in favor of
$\mathbf{E}$, that is, in the frequency domain $\bm{\nabla}\times 
\mathbf{E}=-i\nu\mathbf{B}$ (corresponding to the sign of the Fourier
transform in Eq.~(\ref{fdtee})),
 we can express the desired correlation function in the rest
frame of the blackbody radiation using the fluctuation-dissipation theorem
again:
\be
\langle\mathcal{S}\mathbf{E}(\mathbf{r}_1,t_1)\mathbf{E}(\mathbf{r}_2,t_2)
\rangle=\int_{-\infty}^\infty \frac{d\nu}{2\pi}e^{-i\nu(t_1-t_2)}\Im 
\bm{\Gamma}_R(\mathbf{r}_1,\mathbf{r}_2;\nu)\coth\frac{\beta\nu}2,\label{fdtee}
\ee
where the retarded Green's dyadic $\bm{\Gamma}_R$ is given by Eqs.~(\ref{Gr}) 
and (\ref{gee}),  and $\beta$ is the inverse temperature of the 
blackbody radiation.

Again, let us consider an isotropic particle, so the trace is taken over
the field correlation function in the particle rest frame,
\be
\tr  \langle\mathcal{S}\mathbf{E}'(\mathbf{r}'_1,t'_1)\mathbf{E'}
(\mathbf{r}'_2,t'_2)\rangle=
\int \frac{d\nu}{2\pi}\frac{(d\mathbf{k}_\perp)}{(2\pi)^2}e^{-i\gamma(\nu-k_xv)
(t'_1-t'_2)}e^{i\gamma(k_x-\nu v)(x'_1-x'_2)}f(\nu,k_x)\Im\frac1{2\kappa}\coth
\frac{\beta\nu}2,\label{ecorr}
\ee
where a straightforward bit of algebra yields
\be
f(\nu,k_x)=2\gamma^2v^2\left(\frac\nu{v}- k_x\right)^2.
\ee  Carrying out the frequency Fourier transform of Eq.~(\ref{ecorr})
yields a $\delta$
function setting $\nu=\omega/\gamma+v k_x$, for which $f(\nu,k_x)=2\omega^2$.
The gradient in Eq.~(\ref{fee}) supplies a factor of $ik_x$, since it refers
to the blackbody rest frame. %which requires extracting the imaginary part
%of the polarizability.  
The imaginary part of the Green's dyadic  arises again
from  carrying out the $k_y$ integral over the branch line of 
$1/\kappa$, with the result
\be
(F_x)_{EE}=\frac{i}{8\pi^2\gamma^2}\int_{-\infty}^\infty d\omega\,
\omega^2\alpha(\omega)^*\int dk_x\, k_x 
\sgn\left(\frac{\omega}\gamma+v k_x\right) 
\coth\left[\frac{\beta}2\left(\frac\omega\gamma+v k_x\right)\right],
\label{feefina}
\ee
where the limits on the $k_x$ integration are determined by the
condition $k_x^2<\nu^2$.  Then with $k_x=u\omega$, we have
\be
(F_x)_{EE}=\frac1{8\pi^2\gamma^2}\int_{-\infty}^\infty d\omega\,\omega^4
\Im\alpha(\omega)\int_{-u_-}^{u_+} du\,u\coth\left[\frac{\beta\omega}2
\left(\frac1\gamma+v u\right)\right],\label{feefin}
\ee
where the limits on the $u$ integration are now
 $u_+=\sqrt{\frac{1+v}{1-v}}$ and $u_-=\sqrt{\frac{1-v}{1+v}}$.
Here,  we have used the fact that 
$\alpha(\omega)^*=\alpha(-\omega)$. Thus,
only the imaginary part of $\alpha$ appears because otherwise the 
integrand is odd.
%%%Gerry's addition revised
 Finally, on performing the $u$ integration, we obtain
\bea
(F_x)_{EE}&=& \frac{1}{4\pi^2 v^2 \beta^2\gamma^2}\int_{-\infty}^{\infty} 
d\omega\, \Im \alpha(\omega) \,\omega^2 \left\{\text{Li}_2
\left(e^{-\beta\omega_-}\right) -\text{Li}_2\left(e^{-\beta\omega_+}\right)
-v\beta^2\gamma^2 \omega^2\phantom{\frac{\beta}{2}}\right.\nn\\
&&\qquad\mbox{}+v\beta\left[\omega_{-} \ln\left(2\sinh 
\left(\frac{\beta\omega_-}{2}\right)\right)
+\left.\omega_{+} \ln\left(2\sinh \left(\frac{\beta\omega_+}{2}\right)\right)
\right]\right\},
\eea
where $\omega_-=\omega\sqrt{\frac{1-v}{1+v}}$ and 
$\omega_+=\omega\sqrt{\frac{1+v}{1-v}}$ are the corresponding Doppler-shifted 
frequencies, and $\text{Li}_2$ denotes the dilogarithm function. It is 
immediate that this expression is odd in $v$. That the integrand is even in 
$\omega$, which is already evident from Eq.~(\ref{feefina}), follows from the 
reflection property of the dilogarithm: $\text{Li}_2\left(z^{-1}\right)
=-\text{Li}_2(z)-\frac{\pi^2}{6}-\frac12 \ln^2(-z)$.

%Finally, on performing the $u$ integration, we obtain
%\bea
%(F_x)_{EE}&=& \frac{1}{4\pi^2 v^2 \beta^2\gamma^2}
%\int_{-\infty}^{\infty} d\omega\, 
%\Im \alpha(\omega) \,\omega^2 \left\{-v\gamma^2\beta^2\omega^2+v\beta\omega
%\left[u_{+} \ln\left(2\sinh \left(\frac{u_{+}\beta\omega}{2}\right)\right)
%\right.\right.\nn\\
%&&\qquad +\left.\left.u_{-} \ln\left(2\sinh \left(\frac{u_{-}
%\beta\omega}{2}\right)\right)\right]-\text{Li}_2\left(e^{-u_{+}\beta\omega}
%\right)+\text{Li}_2\left(e^{-u_{-}\beta\omega}\right)\right\},
%\eea
%where $\text{Li}_2$ denotes the dilogarithm function. 

%\noindent  [Please add the below as a subsequent new paragraph.]
%\ms

 An alternative approach is to evaluate the ensemble average of the symmetrized
 field operators {\it directly\/} in the rest frame of the particle. This may 
be achieved by employing the fluctuation-dissipation theorem at the level of 
each $k'_x$ Fourier component, using the corresponding Lorentz-transformed 
inverse temperature from Eq.~(\ref{boostedt}):
\be
\langle \mathcal{S} \bold E'(\bold{r}'_1, t'_1)   \bold E'(\bold{r}'_2, t'_2) 
 \rangle = \int_{-\infty}^{\infty}\frac{d\omega}{2\pi} \,
e^{-i\omega (t'_1-t'_2)} \int\frac{(d\bold{k}'_{\perp})}{(2\pi)^2} \,
e^{i\bold{k}'_{\perp}\bold{\cdot}(\bold{r}'_1-\bold{r}'_2)_{\perp}}  
\Im \bold{g}'_R(z'_1, z'_2; \bold{k}'_{\perp}, \omega)
\coth \left(\frac{\beta\gamma}{2}(\omega+vk'_x)\right).
\ee
The analogue of Eq.~(\ref{gendd}) is then generally
\be
(F_x)_{EE}=\int_{-\infty}^\infty\frac{d\omega}{2\pi}\tr \Im\bm{\alpha}(\omega)
\int\frac{(d\mathbf{k'_\perp})}{(2\pi)^2} (k'_x+v \omega)\Im
\mathbf{g}_R'(0,0;\mathbf{k'_\perp},\omega)\coth
\left(\frac{\beta\gamma}2(\omega+v k'_x)\right).\label{genee}
\ee  Here the imaginary part of the Green's function emerges as before,
so for the isotropic case,
\begin{equation}
(F_x)_{EE}= \frac1{8\pi^2}\int_{-\infty}^{\infty}d\omega\,\Im 
\alpha(\omega)\,\omega^2 \,\text{sgn}(\omega)\int dk'_x \,(k'_x+v\omega) 
\coth\left(\frac{\beta\gamma}{2}(\omega+vk'_x)\right),
\end{equation}
where the limits on the $k'_x$ integration are determined by the condition 
${k'_x}^2< \omega^2$. 
%Evaluation of this integral again results in \thetag{5.15}.
This is equivalent to Eq.~(\ref{feefina}) as seen by making the substitution
$k_x=\gamma(k_x'+v \omega)$.

\subsection{Einstein-Hopf Effect}
The vacuum frictional forces (\ref{fddfin}) and (\ref{feefin}) 
agree with those in Refs.~\cite{vp,v,dedkov,intravaia}, and references therein,
 except for a factor 
of $4\pi$, due to the use of rationalized units in our
case.  The two terms are combined to yield the total vacuum frictional force
\be
F_{\rm tot}=F_{dd}+F_{EE}=-\frac{1}{4\pi^2\gamma^2}\int_0^\infty d\omega \,
\omega^4\Im\alpha(\omega)\int_{-u_-}^{u_+}du\,u\left[\coth\frac{\beta'\omega}2-
\coth\left[\frac{\beta\omega}2\left(\frac1\gamma+vu\right)\right]\right],
\label{ftot}
\ee
which vanishes at zero temperature, and at zero velocity.  We might
expect $\beta'=\gamma\beta$, assuming $\beta$ transforms like the 
time-component of a four-vector.\footnote{ However, Ref.~\cite{vp}
suggests a different, model-dependent, velocity dependence in
equilibrium.  In any case, for low velocities, $\beta'$ should be $\beta$.}  
This is further explored in Appendix 
\ref{appc}.

As pointed out in Ref.~\cite{vp} this result is equivalent to the Einstein-Hopf
effect, which refers to low velocities.  
That force may be written in terms of the blackbody spectral
density $\rho(\omega)$ as (recall that the relation between the 
Heaviside-Lorentz and the more usual Gaussian units for polarizability is 
$\alpha_{\rm HL}=4\pi \alpha_{\rm G}$)
\be
F_x^{\rm EH}=- v \int_0^\infty d\omega\,\omega \Im\alpha(\omega)\left[
\rho(\omega)-\frac\omega3\frac{d}{d\omega}\rho(\omega)\right].\label{feh}
\ee
For the Planck spectrum,
\be
\rho(\omega)=\frac{\omega^3}{2\pi^2}\coth\frac{\beta\omega}2.\label{planck}
\ee
Inserting this into Eq.~(\ref{feh}) we have for the Einstein-Hopf friction
\cite{mkrt} %,lach}
\be
F_x^{\rm EH}=-\frac{2v}{3\pi}\frac1{4\pi}
\int_0^\infty d\omega\,\omega^4\Im\alpha(\omega)
\frac{\beta\omega/2}{\sinh^2\beta\omega/2}.\label{ehfin}
\ee
This is exactly what is obtained from Eq.~(\ref{ftot}) 
if it is expanded for small $v$.

It seems appropriate to conclude this section with a few remarks concerning
numerical magnitudes.  For example, suppose that the moving particle is a small
gold nanosphere of radius  $a$.  The polarizability of such a particle is
\be
\alpha(\omega)=4\pi a^3\frac{\varepsilon(\omega)-1}{\varepsilon(\omega)+2}.
\ee
Let the permittivity be described by the Drude model (\ref{drude}).  Then for
low frequencies, 
(also as seen in Ref.~\cite{vp}, since $\nu=\omega_p^2/(4\pi\sigma)$, 
$\sigma$ being the conductivity)
\be
\Im\alpha(\omega)\approx 4\pi a^3\frac{3\omega\nu}{\omega^2_p}.\label{ns}
\ee
Inserting this into Eq.~(\ref{ehfin}), we obtain
\be
F^{\rm EH}=-v\frac{\nu}a \frac1{21\pi}\frac{(2\pi a/\beta)^6}
{(\omega_p a)^2}\approx -v 10^{-23}\mbox{N},\label{ehvalue}
\ee
where we used values appropriate for gold, $\omega_p=9$ eV, $\nu=0.035$ eV, and
considered room temperature $\beta=40 \mbox{eV}^{-1}$ and a sphere of radius
$a=100$ nm.  This is apparently beyond experimental reach, as we see by
comparing with the usual Casimir-Polder force for a perfectly conducting 
sphere of radius $a$ a distance $d$ above a conducting plate,
\be
F^{\rm CP}=-\frac{3 \alpha_{\rm G}}{2\pi d^5}
=-\frac{3a^3}{2\pi d^5}\sim 10^{-17} 
\mbox{N},
\ee
for $a=100$ nm and $d=1000$ nm, which is already quite small.
(If we had used a radiation reaction model instead, so $\Im\alpha(\omega)
=\frac23\omega^3\alpha_0^2$, as in Ref.~\cite{vp}, a much smaller value
than that in Eq.~(\ref{ehvalue}) would result.)

\begin{figure}
\includegraphics{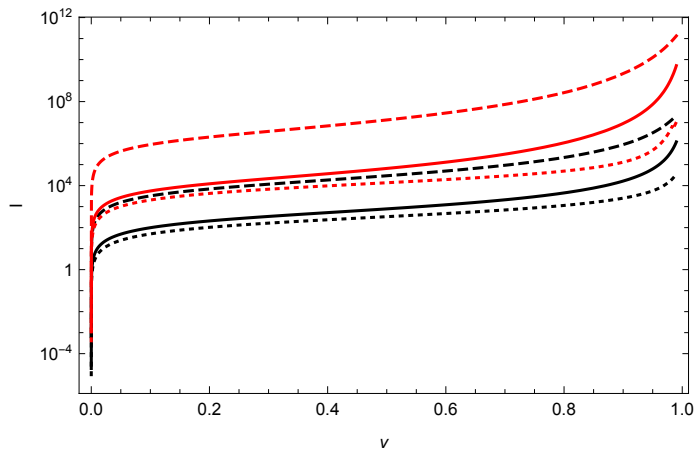}
\caption{\label{qf} The quantum frictional force written in the form of
the dimensionless integral (\ref{ins}) as plotted as a function of the velocity
$v$. The lower set of curves (black) is for $n=1$, and the upper set (red) is
for $n=3$.  In each case, three situations are envisaged: $\beta=\beta'$ 
(solid), $\beta=\beta'/2$ (dashed), and $\beta=2\beta'$ (dotted).}
\end{figure}

 Finally, we note that the integrand in Eq.~(\ref{ftot}) is not
positive definite.  Nevertheless, we would expect the frictional force to 
always be negative, at least when $\Im\alpha(\omega)$ is a power function of
$\omega$.  We illustrate this numerically in Fig.~\ref{qf}, which
plots  the dimensionless integral
\be
I_n(\beta/\beta',v)=\int_0^\infty dx\,x^{4+n}f(x,\beta/\beta',v),
\label{ins}
\ee
where
\be
f(x,\beta/\beta',v)=\int_{-u_-}^{u_+} du\,u\left[\coth\frac{x}2-\coth
\frac{\beta x}{2\beta'}\left(\sqrt{1-v^2}+vu\right)\right];
\ee
here for the ``nanosphere'' model (\ref{ns}) $n=1$, while for the 
radiation-reaction model $n=3$.
To convert these integrals into frictional forces, we have the
following expressions for the two models:
\be
(F_x)^{RR}=-\frac1{6\pi^2\gamma^2}\frac{\alpha_0^2}{\beta^{\prime8}}I_3,
\quad
(F_x)^{NS}=-\frac3\pi\frac\nu{a\gamma^2}\frac{(a/\beta^{\prime})^6}{(\omega_p a)^2}I_1.
\ee
In Appendix \ref{appz} we show that $I_n$ is expressible in closed form,
and always positive.  Further exploration of the sign of
$F$ will appear elsewhere.

\section{Conclusions}
\label{concl} 
What is remarkable about the straightforward calculations sketched in this 
paper is that starting simply from Maxwell's equations, and the corresponding
Lorentz force law, familiar in elementary physics, one {\em deduces\/}
the formula for the energy emitted by dipole radiation, without ever
invoking radiation fields or the concept of radiation reaction.
This analysis may appear similar to the discussion of friction due to 
thermal \cite{eh} or zero-point \cite{thb} fluctuations, in which
a dipole, including radiation reaction, is coupled to fluctuating
electromagnetic fields.  In the latter case, of course, no force on
the dipole is found at zero temperature
\cite{pm,milonni}.  The classical theory does not involve  fluctuations.

To demonstrate that our considerations are sensible, 
we recover known results for
the spectral distribution of the Vavilov-\v Cerenkov radiation emitted
by a superluminal electric or magnetic dipole.  In the latter case,
we confirm, in accordance with  Refs.~\cite{frank84,leonhardt19} that there
is a discontinuity in the radiation at threshold, $v n(\omega)=1$, 
for polarization of the
magnetic dipole perpendicular to the direction of motion.

We then present  a discussion of the classical electromagnetic friction 
experienced by a time-independent
 electric or  magnetic dipole moving parallel to  
an imperfectly conducting surface.  We derive general
formulas valid for all velocities and polarizations. Explicit results are
given for the low velocity region, $v\ll a\nu$ where $\nu$ is the damping
parameter, proportional to the resistivity of the conductor, and $a$ is
the distance of the dipole from the surface of the conductor.  This is likely
the regime in which this friction might be more accessible to experiment.
% observed in the laboratory.

We finally show how by minimal use of the fluctuation-dissipation theorem
applied to both dipoles and fields we can recover the Einstein-Hopf effect,
which, for arbitrary velocities, yields the quantum blackbody friction
in vacuum for a polarizable particle found by Dedkov and Kyasov \cite{dedkov}
and Volokitin and Persson \cite{vp}.
In  Appendix \ref{appb} we show that the same considerations give rise to the
usual Casimir-Polder interaction with an arbitrary body.
In subsequent publications, we will further explore quantum friction 
in the vicinity of other bodies.

\acknowledgments{We thank our collaborators Prachi Parashar and Steve
Fulling for many insightful comments on this work.  This research
was supported in part by the US National Science Foundation,
grant number 1707511.}

\appendix
\section{Lorentz Transformation of Charge and Current Densities}
\label{app0}
Since it appears a bit subtle for a time-dependent dipole, we sketch
the inference of the charge and current density for a moving dipole given
in Eq.~(\ref{chargeandcurrent}) from that for a stationary dipole,
\be
\rho'(\mathbf{r'},t')=-\mathbf{d}'(t')\cdot\bm{\nabla}'\delta(\mathbf{r'}),
\quad
\mathbf{j}'(\mathbf{r'},t')=\mathbf{\dot d}'(t')\delta(\mathbf{r'}).
\ee
Recall that unprimed quantities refer to the frame where the particle has
velocity $\mathbf{v}=\mathbf{\hat x}v$, while primed quantities refer to the
particle's rest frame.  The dot always means derivative with respect to the
argument. The coordinates in the two frames are related by
\be
x=\gamma(x'+vt'),\quad t=\gamma(t'+vx'),\quad y=y',\quad z=z'.
\ee
Now $(\rho,\mathbf{j})$ constitutes a four-vector, so
\bea
\rho(\mathbf{r},t)&=&\gamma[\rho'(\mathbf{r}',t')+v j'_x(\mathbf{r}',t')]\nn\\
&=&\gamma[-d'_x(\gamma(t-vx))\gamma(\partial_x+v\partial_t)-\mathbf{d}'_\perp
(\gamma(t-vx))\cdot\bm{\nabla}_\perp+v\dot{d}'_x(\gamma(t-vx))]
\delta(\gamma(x-vt))\delta(\mathbf{x}_\perp),
\eea
where the $\perp$ sign signifies the $y$, $z$ directions. %and the dot
%means derivative with respect to the argument.  
We use the identity
\be
f(x)\partial_x \delta(x)=f(0)\partial_x \delta(x)-\delta(x)\partial_x f(0),
\label{identity}
\ee to write the above as
\bea
\rho(\mathbf{r},t)&=&
\bigg[-\frac1\gamma d'_x\left(\frac{t}\gamma\right)\partial_x \delta(x-vt)
+\frac1\gamma\delta(x-vt)\partial_x d'_x(\gamma(t-vx))
-\mathbf{d}'_\perp\left(\frac{t}\gamma\right)\cdot\bm{\nabla}_\perp\delta(x-vt)
\nn\\
&&\quad\mbox{}+v\dot{d}'_x(\gamma(t-vx))\delta(x-vt)\bigg]\delta(y)\delta(z)
=-\mathbf{d}(t)\cdot\bm{\nabla}\delta(x-vt)\delta(y)\delta(z),
\eea
which uses the transformation properties for the dipole moments, 
Eq.~(\ref{ltofdip}), and notes that the second and fourth terms in the
square brackets cancel.

The $x$-component of the current, 
\bea
j_x(\mathbf{r},t)&=&\gamma[j'_x(\mathbf{r}',t')+v\rho'(\mathbf{r}',t')]\nn\\
&=&[\dot d'_x(\gamma(t-vx))-vd'_x(\gamma(t-vx))\gamma(\partial_x+v\partial_t)
-v \mathbf{d}'_\perp(\gamma(t-vx))\cdot\bm{\nabla}_\perp]\delta(x-vt)\delta(y)
\delta(z),
\eea
%Since the dot signifies derivative with respect to the argument,  in terms
%of the moving frame coordinates it equals $\gamma(\partial_t+v\partial_x)$.
becomes, using the identity (\ref{identity}), 
\be
j_x(\mathbf{r},t)=\left\{\dot d'_x(\gamma(t-vx))-\frac{v}\gamma
\left[d'_x\left(\frac{t}\gamma\right)\partial_x
+v\partial_t d'_x(\gamma(t-vx))\right]
-v\mathbf{d}'_\perp\left(\frac{t}\gamma\right)
\cdot\bm{\nabla}_\perp\right\}\delta(x-vt)\delta(y)
\delta(z).\label{jxt}
\ee
%The first term here is is evaluated by Taylor expanding the argument around
%$x=vt$:
%\be
%d_x'(\gamma(t-vx))=d_x'\left(\frac{t}\gamma\right)
%+(x-vt)\partial_x d'_x(\gamma(t-vx))
%\big|_{x-vt},
%\ee
The first and third terms here combine to give $\frac1{\gamma^2}\dot d'_x
(\gamma(t-vx))$
%the second term here resulting in the cancellation of
%the third term in Eq.~(\ref{jxt}), 
so when we use the transformation
properties (\ref{ltofdip}) we obtain the expected result,
\be
j_x(\mathbf{r},t)=[-v \mathbf{d}(t)\cdot\bm{\nabla}+\dot d_x(t)]\delta(x-vt)\delta(y)
\delta(z).
\ee
More immediately, the $y$-component of the current is transformed to
\be
j_y(\mathbf{r},t)=\dot d_y(t)\delta(x-vt)\delta(y)\delta(z).
\ee
Thus the charge and current densities due to an electric dipole in the moving
frame of the particle, Eq.~(\ref{chargeandcurrent}), 
are derived from the rest-frame form.
A similar argument applies for the charge and current densities  
due to a magnetic dipole, Eq.~(\ref{magchcur}).

\section{Relation between Green's functions}
\label{appa}
In Sec.~\ref{sec8} we used the symmetrized correlation function of the fields
given in terms of the imaginary part of the retarded Green function by the
fluctuation-dissipation theorem, with $n_\nu=(e^{\beta\nu}-1)^{-1}$:
\be
\langle \mathcal{S} \mathbf{E}(\mathbf{r}_1,t_1)\mathbf{E}(\mathbf{r}_2,t_2)
\rangle=\int_{-\infty}^\infty \frac{d\nu}{2\pi}\Im\bm{\Gamma}_R
(\mathbf{r,r'};\nu)(2n_\nu+1)e^{-i\nu(t_1-t_2)}.
\ee
In quantum field theory we usually use the time-ordered product,
\be
\langle \mathcal{T}\mathbf{E}(\mathbf{r}_1,t_1)\mathbf{E}(\mathbf{r}_2,t_2)
\rangle= \int_{-\infty}^\infty \frac{d\nu}\pi\Im\bm{\Gamma}_R(\mathbf{r}_1,
\mathbf{r}_2;\nu)(n_\nu+1)e^{-i\nu|t_1-t_2|}.
\ee
The Fourier transform of this gives the thermal Green's function
\be
\bm{\Gamma}_{+\beta}(\mathbf{r}_1,\mathbf{r}_2;\omega)
=i\langle\mathcal{T}\mathbf{EE}
\rangle(\mathbf{r}_1,\mathbf{r}_2;\omega)= \frac2\pi
\int_{-\infty}^\infty d\nu\,\nu (n_\nu+1)
\frac{\Im\bm{\Gamma}_R(\mathbf{r}_1,\mathbf{r}_2;\nu)}{(\nu-i\epsilon)^2
-\omega^2}.
\ee
This is to be contrasted with the representation for the retarded Green's 
function, which has the form of a Kramers-Kronig relation,
\be
\bm{\Gamma}_R(\mathbf{r}_1,\mathbf{r}_2;\omega)=\frac2\pi\int_0^\infty
d\nu\,\nu\frac{\Im\bm{\Gamma}_R(\mathbf{r}_1,\mathbf{r}_2;\nu)}{\nu^2-(\omega
+i\epsilon)^2},
\ee
which has no temperature dependence.  The relation between the imaginary
parts is
\begin{subequations}
\be
\Im\bm{\Gamma}_{+\beta}(\mathbf{r}_1,\mathbf{r}_2;\omega)=\Im\bm{\Gamma}_R
(\mathbf{r}_1,\mathbf{r}_2;\omega)(2n_\omega+1), \quad 2n_\omega+1=\coth
\frac{\beta\omega}2,\label{ims}
\ee
while the real parts are the same,
\be
\Re\bm{\Gamma}_{+\beta}(\mathbf{r}_1,\mathbf{r}_2;\omega)=\Re\bm{\Gamma}_R
(\mathbf{r}_1,\mathbf{r}_2;\omega).\label{res}
\ee
\end{subequations}

It is most usual to evaluate Casimir energies by integrating over 
Euclidean frequencies, $\omega\to i\zeta$.  In that case, the Green's functions
are identical:
\be
\bm{\Gamma}_{+\beta}(i\zeta)=
\frac2\pi\int_{-\infty}^\infty d\nu\,\nu
\frac{\Im\bm{\Gamma}_R(\nu)}{\nu^2+\zeta^2}
(n_\nu+1)
=\frac2\pi\int_0^\infty d\nu\,\nu
\frac{\Im\bm{\Gamma}_R(\nu)}{\nu^2+\zeta^2}=\Gamma_R(i\zeta).
\ee

\section{Static Casimir-Polder Energy}
\label{appb}
Consider the static Casimir-Polder interaction
between some (unspecified) background object and a polarizable atom.  The
contribution to the interaction free energy due to field fluctuations is
\be
\mathcal{F}_{EE}=-\frac12\tr\int_{-\infty}^\infty \frac{d\omega}{2\pi}
\bm{\alpha}(\omega)^*\langle
\mathcal{S}\mathbf{EE}\rangle(\omega)=-\frac12\int_{-\infty}^\infty
\frac{d\omega}{2\pi}\tr\Re\bm{\alpha}(\omega)\Im\bm{\Gamma}_R(\omega)
\coth\frac{\beta\omega}2,
\ee
since the imaginary part of the polarizability is odd.
The second contribution to the energy comes from 
the dipole fluctuations, that is
\be
\mathcal{F}_{dd}
=-\frac1{2T}\Tr\mathbf{j}^*\frac1{\omega^2}\bm{\Gamma}\mathbf{j},
\ee
which follows from Eq.~(\ref{currentcurrent}).  This immediately leads to,
upon use of the current for a stationary dipole $\mathbf{j}(\mathbf{r},\omega)
=-i\omega\mathbf{d}(\omega)\delta(\mathbf{r})$,
%the $v=0$ form of Eq.~(\ref{fet}),
\be
\mathcal{F}_{dd}=-\frac1{2T}\int_{-\infty}^\infty \frac{d\omega}{2\pi}
\left\langle\mathcal{S}
\mathbf{d}(\omega)^*\cdot\bm{\Gamma}_R(\omega)\cdot\mathbf{d}(\omega)
\right\rangle
=-\frac12\int_{-\infty}^\infty \frac{d\omega}{2\pi}\coth\frac{\beta\omega}2\tr
\Im\bm{\alpha}(\omega)\cdot\Re\bm{\Gamma}_R(\omega).
\ee
Thus the total free energy is just as expected:
\be
\mathcal{F}_{CP}=\mathcal{F}_{dd}+\mathcal{F}_{EE}
=-\frac12\int_{-\infty}^\infty\frac{d\omega}{2\pi}
\coth\frac{\beta\omega}2\Im\tr\left[\bm{\alpha}(\omega)\bm{\Gamma}_R(\omega)
\right].\label{fcp}
\ee
This result could be recaptured using the time-ordered polarizability and the 
thermal Green's function by use of Eq.~(\ref{ims}) and Eq.~(\ref{res}), 
\be
\mathcal{F}_{CP}=-\frac12 \int_{-\infty}^\infty\frac{d\omega}{2\pi} 
\Im\tr [\bm{\alpha}_{+\beta}(\omega)\bm{\Gamma}_{+\beta}(\omega)],
\ee
where
\be
\bm{\alpha}_{+\beta}(\omega)=i\langle\mathcal{T}\mathbf{dd}\rangle(\omega)=
\frac2\pi\int_{-\infty}^\infty d\nu\frac{\nu(n_\nu+1)\Im\bm{\alpha}(\nu)}
{(\nu-i\epsilon)^2-\omega^2}.
\ee

The form in Eq.~(\ref{fcp}) is perhaps more familiarly expressed
in terms of  Euclidean frequencies.  Since the retarded
Green's function and the polarizability
have no singularities in the upper half $\omega$ plane,
the contour of integration can be distorted to one encircling the positive
imaginary axis, and then accounting for the poles of the cotangent along
that axis, we obtain
\be
\mathcal{F}_{CP}
=-\frac1\beta\sum_{n=0}^\infty{}'\tr\bm{\alpha}(i\zeta_n)\bm{\Gamma}(i\zeta_n),
\ee
in terms of the Matsubara frequency $\zeta_n=2\pi n/\beta$.
(The prime on the summation sign means that the $n=0$ term is counted with 
half weight.)

\section{Transformation of blackbody spectral density}
\label{appc}
It was shown by Ford and O'Connell \cite{ford} that the spectral density
(\ref{planck}) becomes in a frame moving with velocity $\mathbf{v}$
\be
\rho'(\omega',\mathbf{k}')=\frac{\omega^{\prime 3}}{2\pi^2}\coth\left[
\frac{\beta}2\gamma\omega'(1+\mathbf{\hat k'}\cdot \mathbf{v})\right].
\ee
Here $\beta$ is the inverse temperature in the blackbody rest frame.
This is, in fact, just what we would expect if $\beta$ is thought to be
the time component of four-vector, $\beta=\beta_0$, $\bm{\beta}=0$.
So we could write
\be 
\rho(\omega)=\rho(\omega,\mathbf{k})=\frac{\omega^3}{2\pi^2}\coth
\frac12\beta_\mu k^\mu,\quad k^\mu=(\omega,\mathbf{k}).
\ee
Indeed this would become
\be
\rho'(\omega',\mathbf{k}')=\frac{\omega^{\prime3}}{2\pi^2}
\coth\frac12\beta'_\mu k^{\prime\mu},
\ee
where for a boost in the $x$ direction,
\be
\beta'_\mu k^{\prime\mu}=\gamma\beta\omega'+\gamma v\beta k_x'
=\gamma\beta\omega'\left(1+v\frac{k'_x}{k'}\right), \quad k'=\omega'.
\label{boostedt}
\ee
So the result of Ford and O'Connell corresponds to the expected transformation
of $\beta_\mu$.
 
\section{Evaluation of the integral $I_n$}
\label{appz}
To evaluate the integral $I_n$ we carry out the $x$ integration first,
with the result
\be
I_n=2\Gamma(5+n)\zeta(5+n)\int_{-u_-}^{u_+} du\,u\left[1-\left(\frac\beta
{\beta'}\left[\sqrt{1-v^2}+vu\right]\right)^{-5-n}\right],
\ee
and then integrating by parts on $u$ we obtain
\bea
I_n&=&2\Gamma(5+n)\zeta(5+n)\bigg\{2v\gamma^2+\frac1{(3+n)(4+n)}
\frac{\gamma^{3+n}}{v^2}\left(\frac{\beta'}\beta\right)^{3+n}\nn\\
&&\qquad\times\left[
(1-v)^{3+n}-(1+v)^{3+n}+(3+n)v\left((1-v)^{3+n}+(1+v)^{3+n}\right)\right]
\bigg\}.
\eea
This is positive for all $v$, if $n>-3$,
 as may be seen by expanding the quantity in
square brackets in powers of $v$: the coefficient of $v^{2k+1}$ is 
$2k(4+n)$.


\begin{thebibliography}{9}

%\cite{Milton:2015aba}
\bibitem{Milton:2015aba} 
  K.~A.~Milton, J.~S.~H\o ¸ye and I.~Brevik,
  ``The reality of Casimir friction,''
  Symmetry {\bf 8}, no. 5, 29 (2016)
  doi:10.3390/sym8050029
  [arXiv:1508.00626 [quant-ph]].
  %%CITATION = doi:10.3390/sym8050029;%%
  %10 citations counted in INSPIRE as of 19 Feb 2020

\bibitem{clfriction}
K. A. Milton, Y. Li, X. Guo, and G. Kennedy, 
``Electrodynamic friction of a charged particle passing a conducting plate,'' 
 Phys.\ Rev. Research {\bf 2}, 023114 (2020) [arXiv:1911.06369]

\bibitem{eh} A. Einstein and L. Hopf, ``Statistische Untersuchung
der Bewegung eines Resonators in einem Strahlungsfeld,'' Ann. Phys. 
(Leipzig) {\bf 336}, 1105--1115 (1910).

\bibitem{thb} T. H. Boyer, ``Derivation of the blackbody radiation
spectrum without quantum assumptions,'' Phys.\ Rev.\ {\bf 182}, 1374
(1969).

\bibitem{frank} I. M. Frank, ``Doppler effect in a refractive medium,''
Izv.\ Akad.\ Nauk SSSR {\bf 6}, 3 (1942).



\bibitem{frank84} I. M. Frank, ``Vavilov-Cherenkov radiation for electric
and magnetic multipoles,'' Usp.\ Fiz.\ Nauk {\bf 44}, 251--275 (1984) [Sov.\
Phys.\ Usp.\ {\bf 27}, 772 (1984)].

\bibitem{dedkov} G. V. Dedkov and A. A. Kyasov, ``Radiation of a neutral
polarizable particle moving uniformly through a thermal radiation field,''
Phys.\ Scr.\ {\bf 89}, 105501 (2014).


\bibitem{vp} A. I. Volokitin and B. N. J. Persson, {\it Electromagnetic
Fluctuations at the Nanoscale} (Springer, Berlin, 2017), Chap.~8.

\bibitem{js} J. Schwinger, {\it Particles, Sources, and Fields} 
(Addison-Wesley, Reading, 1970).

\bibitem{ce} J. Schwinger, L. L. DeRaad, Jr., K. A. Milton, and W.-y. Tsai,
{\it Classical Electrodynamics} (Perseus/Westview, Taylor and Francis,
1998).

\bibitem{cherenkov}
P. A. Cherenkov, ``Visible emission of clean liquids by action of 
$\gamma$³ radiation,'' Dokl.\ Akad.\ Nauk SSSR {\bf2} 451 (1934). 

\bibitem{tamm-frank}
I. E. Tamm and I. M. Frank, ``Coherent radiation of fast electrons in a 
medium,'' Dokl.\ Akad.\ Nauk SSSR {\bf14},  107 (1937).



\bibitem{leonhardt19} U. Leonhardt and Y. Rosenberg, ``Cherenkov radiation
of light bullets,'' Phys.\ Rev.\ A {\bf 100}, 063802 (2019).

\bibitem{ford2} G. W. Ford, ``The fluctuation-dissipation theorem,''
Cont.\ Phys.\ {\bf 58}, 244--252 (2017).

\bibitem{kubo} R. Kubo, ``The fluctuation-dissipation theorem,'' 
Rep.\ Prog.\ Phys.\ {\bf 29}, 255--284 (1966).

\bibitem{v} A. I. Volokitin, ``Friction force at the motion of a small 
relativistic particle with respect to blackbody radiation,'' Pis'ma Zh.\
Eksp.\ Fiz.\ {\bf 101}, 479 (2015) [JETP Lett.\ {\bf 101}, 427 (2015)].


\bibitem{intravaia} F. Intravaia, C. Henkel, and M. Antezza, 
``Fluctuation-induced forces between atoms and surfaces: The Casimir-Polder
interaction,''
in {\it Casimir Physics} ({\it Lecture Notes in Physics}, vol.\ 834), ed.
D. A. R. Dalvit, P. W. Milonni, D. Roberts, and F. Da Rosa (Springer, Berlin,
2011), pp.~345--391.

\bibitem{mkrt} V. Mkrtchian, V. A. Parsegian, R. Podgornik, and W. M. Saslow,
``Universal thermal radiation drag on neutral objects,'' Phys.\ Rev.\ Lett.\
{\bf 91}, 220801 (2003).

%{\color{red}
%\bibitem{lach} G. {\L}ach, M. DeKieviet, and U. D. Jentschura,
%``Einstein-Hopf drag, Doppler shift of thermal radiation and blackbody 
%friction: A unifying perspective on an intriguing physical effect,''
%Cent.\ Eur.\ J. Phys.\ {\bf 10}, 763--767 (2012).
%}

\bibitem{pm} P. W. Milonni, ``Quantum Mechanics of the Einstein-Hopf Model,''
Am.\ J. Phys.\ {\bf 49}, 177-184 (1981).

\bibitem{milonni} P. W. Milonni, {\it The Quantum Vacuum: An Introduction
to Quantum Electrodynamics} (Academic Press, Boston, 1994).
  
\bibitem{ford} G. W. Ford and R. F. O'Connell, ``Lorentz transformation
of blackbody radiation,'' Phys.\ Rev.\ E, {\bf 88}, 044101 (2013). 

\end{thebibliography}
\end{document}